\documentclass[conference]{IEEEtran}
\IEEEoverridecommandlockouts

\def\BibTeX{{\rm B\kern-.05em{\sc i\kern-.025em b}\kern-.08em
    T\kern-.1667em\lower.7ex\hbox{E}\kern-.125emX}}

\usepackage{graphicx}
\usepackage{balance} 
\usepackage{amsmath,amssymb,amsfonts}
\usepackage{textcomp}
\usepackage{latexsym}
\usepackage{color}
\usepackage{epsfig}
\usepackage{xspace}
\usepackage{pbox}
\usepackage{caption}
\usepackage{epsfig}
\usepackage{url}
\usepackage{mathrsfs}
\usepackage{booktabs}
\usepackage{adjustbox}
\usepackage{array}
\usepackage{multirow}
\usepackage{empheq}
\usepackage{tcolorbox}
\usepackage{stfloats}
\usepackage{enumitem}
\usepackage[colorinlistoftodos]{todonotes}
\usepackage{algorithm}
\usepackage{algpseudocode}
\usepackage{listings}
\usepackage{fbox}
\usepackage{xcolor}
\usepackage{cite}
\usepackage{titlesec}
\usepackage{capt-of}
\usepackage{booktabs}
\usepackage[table]{xcolor}
\usepackage{makecell}
\usepackage{hyperref}

\newcommand{\stitle}[1]{\vspace{.5ex}\noindent{\bf #1}}
\newcommand{\eetitle}[1]{\vspace{0.2ex}\noindent{\underline{\em #1}}}

\newcommand{\sstab}{\rule{0pt}{8pt}\\[-2.4ex]}

\newcommand{\ie}{\emph{i.e.,}\xspace}

\newcommand{\wrt}{\emph{w.r.t.}\xspace}

\newcounter{example}
\renewcommand{\theexample}{\arabic{example}}
\newenvironment{example}{
        \vspace{1.5ex}
        \refstepcounter{example}
        {\noindent\bf Example \theexample:}}

\newcommand{\bi}{\begin{itemize}}
\newcommand{\ei}{\end{itemize}}

\newcommand{\eat}[1]{}

\newcommand{\one}[1]{{\color{black}{#1}}}

\newcommand{\three}[1]{{\color{black}{#1}}}

\newcommand{\llm}{\kw{LLM}} 
\newcommand{\llms}{\kw{LLMs}}

\newcommand{\kw}[1]{{\ensuremath{\mathsf{#1}}}\xspace}

\newcommand{\free}{\kw{Freebase}}

\newcommand{\webqsp}{\kw{WebQSP}} 
\newcommand{\grail}{\kw{GrailQA}} 
\newcommand{\gq}{\kw{GraphQ}}

\newcommand{\cc}{\kw{UniQGen}}
\newcommand{\uqg}{\kw{UniQGen}}

\newcommand{\qc}{\kw{QChase}}
\newcommand{\qbc}{\kw{QBackchase}}
\newcommand{\blone}{\kw{Prompt}-\kw{Only}}

\newcommand{\pangu}{\kw{Pangu}} 
\newcommand{\arcane}{\kw{ArcaneQA}} 
\newcommand{\uni}{\kw{Universal}} 
\newcommand{\mini}{\kw{Minimal}} 
\newcommand{\all}{\kw{Optimal}} 

\newcommand{\exact}{\kw{Exact Match}} 
\newcommand{\f}{\kw{F1}}

\newcommand{\ctable}{\kw{CTable}}
\newcommand{\C}{{\mathcal C}}
\newcommand{\e}{{\varepsilon}}
\newcommand{\g}{{\gamma}}
\newcommand{\agent}{\kw{Agent}}
\newcommand{\con}{\kw{CTable} \kw{Constructor}}
\newcommand{\gen}{\kw{Query} \kw{Generator}}
\newcommand{\eva}{\kw{Evaluator}}

\newcommand{\lightmidrule}{\arrayrulecolor{gray!85}\midrule\arrayrulecolor{black}}

\newcommand{\revise}[1]{\textcolor{black}{#1}}

\newcounter{metactr}

\begin{document}

\title{Graph Query Generation with Constraint-guided Large Language Agents} 

\author{
    \IEEEauthorblockN{
        Mengying Wang$^{*1}$,
        Nicolaas Jedema$^{2}$,
        Rahul Pandey$^{2}$,
        RaviKiran Krishnan$^{*3}$,
        Jens Lehmann$^{2,4}$,
        Yinghui Wu$^{1}$
    }

    \IEEEauthorblockA{
    $^{1}$Case Western Reserve University; 
    $^{2}$Amazon;
    $^{3}$Meta;
    $^{4}$Technische Universität Dresden\\
    Email: \{mxw767, yxw1650\}@case.edu; 
        \{jedema, panrahu, jlehmnn\}@amazon.com; 
        ravi.krishnan@outlook.com}

    \thanks{$^{*}$ Work done while at Amazon AGI.}
}

\makeatletter
\def\@IEEEaftertitletext{\vspace{-2\baselineskip}}
\def\@IEEEafterauthortext{\vspace{-1\baselineskip}}
\makeatother

\maketitle

\begin{abstract}
Knowledge Graph Question Answering (KGQA) has advanced through structured query generation, yet most efforts target RDF/SPARQL, leaving Cypher and property graphs underexplored, despite increasing demand for unified KGQA in industry settings.
We propose \uqg, a novel constraint‑based framework that employs \llm agents to dynamically extract and refine representative 
graph query clauses into executable, intent-aligned graph queries across query languages.
The foundation of our method is a variant of 
 Chase \& Backchase, a family of algorithms for query optimization and reformulation. 
 We extend Chase \& Backchase with a dynamic reasoning process over 
 query constraints that also interact with \llms for 
 query quality estimation. 
With a Cypher-supported Freebase graph deployed on Amazon Neptune,
we extensively evaluate our approach on popular KGQA
benchmarks (\gq, \grail, and \webqsp). 
We demonstrate 
that \cc outperforms state-of-the-art graph 
query generation techniques 
in both accuracy and efficiency, with F1 gains of 31.6$\%$ on \gq\ and 4.9$\%$ on \grail. 
Unlike prior methods, our framework does not require fine-tuning for schema matching, making it more extensible to schema-less graphs and 
semantics in query workloads, and is more suitable for enterprise-grade KGQA.
We release Cypher outputs and a Neptune-ready Freebase snapshot to support reproducible, cross-language KGQA research.

\end{abstract}


\section{Introduction}
\label{sec:intro}

Graph querying have been routinely applied 
to support various tasks such as knowledge-graph question answering (KGQA). 
Current solutions have been deployed for two isolated 
scenarios: the Resource Description Framework (RDF) model, queried with SPARQL; 
or the Labeled Property Graph (LPG) model, queried by Cypher/Gremlin.
This often leads to a ``graph–model lock‑in'' effect~\cite{lassila2022onegraph}:
once an organization commits to a paradigm, that choice propagates to the query
language, toolchain, and developer expertise, making later migration costly and risky. In response, recent efforts advocate a unified 
interface that can cope with 
different data models, 
including RDF, RDF$^\star$, and LPG~\cite{angles2022multilayer,gelling2023bridging}.
For example, Amazon Neptune’s OneGraph~\cite{lassila2022onegraph}
advocates a unified store exposed through multiple query languages. Recent
openCypher‑over‑RDF demonstrations~\cite{schmidt2024opencypher} offer a 
storage‑level bridge that allows Cypher syntax to query RDF data. 

To enable a unified interface for KGQA, it also calls for a 
``query‑level'' effort: a query generation mechanism 
that can automatically 
convert user intent
to mainstream graph query classes
such as Cypher or SPARQL, 
mitigating ``language lock‑in'' effect. 
Most public KGQA benchmarks support  
a single query class, such as 
RDF/SPARQL‑centric (e.g., GrailQA~\cite{gu2021beyond}, WebQSP~\cite{yih2016value}). 
This leaves a gap between academic solutions and industrial practice, 
despite languages such as 
Cypher underpins many production stacks including 
popular graph databases such as Neo4j~\cite{guia2017graph}.  

\begin{figure}[t!]
  \centering
  \includegraphics[width=0.46\textwidth]{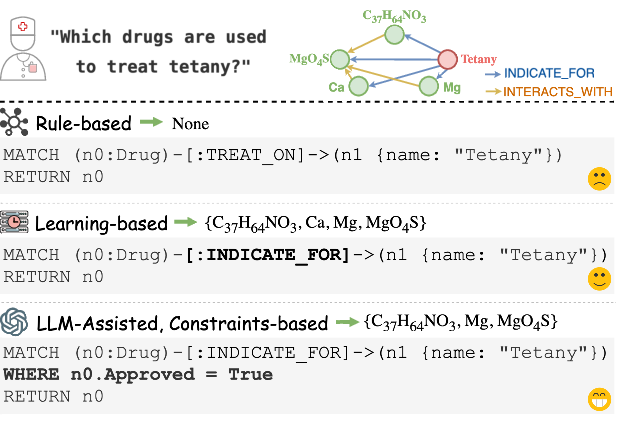}
  \caption{\small
  Given a query: \emph{``Which drugs are used to treat tetany?''}
  \textbf{Rule‑based} methods assume a non-existent relation and returns None;
  \textbf{Learning-based} methods reach the disease but over-includes \texttt{Ca}, failing to address pragmatic constraints; \textbf{LLM‑assisted, constraint-based} methods generation captures implicit intent 
  and respects ontology, yielding clinically
  appropriate answers.}
  \label{fig:motivation}
  \vspace{-3ex}
\end{figure}

Meanwhile, \llm-based methods become attractive for graph query generation due to their strong capabilities in both contextual understanding and code generation~\cite{drozdov2022compositional, li2024flexkbqaflexiblellmpoweredframework, ye2021rng,agarwal2024bring}, achieving state‑of‑the‑art results on multiple SPARQL benchmarks~\cite{gu2023don, shu2022tiaramultigrainedretrievalrobust}.
Nevertheless, in industrial
settings they exhibit well‑known challenges:  
(i) \emph{Hallucination/misalignment}: syntactically valid queries do not
match the KG’s schema/topology (top row of Fig.~\ref{fig:motivation});
(ii) \emph{Missing pragmatic constraints}: queries fails to 
return answers for desired search intent, 
such as ``approved drug'' (middle row of Fig.~\ref{fig:motivation});
(iii) \emph{High maintenance cost}: 
heavy rely on schema‑specific fine‑tuning makes
updates and migration of solutions brittle and costly, especially 
for large ontologies and KGs (e.g., Freebase/Wikidata); 
(iv) \emph{Lack of unified, principled solution}: training and maintaining separate models per query language is costly, and cold‑start for understudied languages like Cypher hinders cross‑language adoption and fair comparison.

In response, we propose \uqg, a schemaless, training‑free solution that \revise{turns Chase and Backchase algorithm into a budgeted, cross-language query generation pipeline by introducing an online-constructed, uncertainty-scored constraint IR (\ctable) and an LLM-assisted pluggable renderer, rather than relying on offline dependencies or training; it offers relative guarantees \wrt a fixed reference answer set $A$ (user- or oracle-provided) by accepting candidates only after runtime validations.}


\begin{example}
\label{exa:motivation}
As shown in Fig.~\ref{fig:motivation}, for query ``Which drugs are used to treat tetany?'', the rule-based methods directly extract semantic information from the natural language query, which assumes a direct \texttt{TREAT\_ON} edge returns no result when the KG encodes indications as \texttt{INDICATE\_FOR}; learning‑based methods capture the schema through costly training processes, while they may identify \texttt{INDICATE\_FOR}, still output an over‑inclusive set because it lacks implicit constraints (e.g., approval status).
In contrast, the constraint-based LLM-assisted method constructs a compact constraint table that (i) fixes the
topology with typed edges, and (ii) makes the implicit intent explicit
(\texttt{Approved=True}); a Chase \& Backchase loop then prunes irrelevant atoms and
returns a precise, executable query.
\end{example}

\stitle{Contributions. }
We summarize our contributions as follows:
\begin{itemize}
\item \textbf{Deployment‑friendly unified pipeline (\uqg). } 
We design an 
autonomous pipeline 
that unifies constraint extraction, \ctable\ construction, and a Chase \& Backchase‑based query refinement process to generate
grammatically‑correct, semantically reasonable queries with low generation latency, for schema-less KGQA. 

\item \textbf{LLM‑assisted constraint-based reformulation method. } We adapt Chase \& Backchase to graph query generation: \llm agents explore and rank candidate
constraints via beam search, while Chase \& Backchase enforces relative completeness,
soundness, and minimality. 
It extends to other query languages with modest effort.

\item \textbf{Experiments and Resources. } 
We build on Amazon Neptune with a deployed Freebase endpoint that supports Cypher execution for fair cross‑language comparison with SPARQL‑only SOTA.
We release a well‑curated Freebase snapshot with a one‑step deployment recipe on Amazon Neptune to enable rapid, large‑scale evaluation~\cite{shirvani2023comprehensive}, 
along with gold Cypher pairs aligned with popular SPARQL benchmarks to facilitate future research.
\end{itemize}

\stitle{Related Works} We categorize the related works as follows:

\eetitle{Graph Query Generation}. 
Recent efforts most leverage \llms
to improve the quality of SPARQL query generation~\cite{ rony2022sgpt, nie2024code}. 
However, existing approaches often require large training data and are tied to knowledge graph schemas, limiting their adaptability~\cite{diallo2024comprehensive}.
Cypher query generation for property graphs has received less attention~\cite{nie2022graphq}. 
Our approach addresses unified query generation in a schemaless and training-free manner, making it adaptable to a wide range of 
querying scenarios, and refined by Chase \& Backchase  algorithm~\cite{deutsch2006query} with quality guarantees. This is not 
addressed in prior methods. 

\eetitle{LLM-enhanced KGQA}. 
\llms have been used to structure NL inputs, perform multi‑step reasoning, and
produce executable queries in KGQA area~\cite{sen2023knowledge, le2024graphlingo}. 
These systems typically emphasize semantic parsing quality on SPARQL‑centric
benchmarks.
Inspired by~\cite{jin2024graph}, \uqg employs \llm agents not just for static query generation but also for dynamically extracting constraints from NL queries and interacting with KGs. This allows for more interactive, contextually aware, and refined query outputs, ensuring high-quality results.

\eetitle{Unified KGQA.}
There is growing demand, especially in industry, for KGQA systems that operate seamlessly across RDF/SPARQL and LPG/Cypher stacks, yet most efforts focus on data‑model~\cite{angles2022multilayer, gelling2023bridging} or engine‑level~\cite{lassila2022onegraph} unification, 
but lack a light‑weight, modular way to transfer NL intent into executable queries across languages.
\revise{
GraphQ IR~\cite{nie2022graphq} proposes an intermediate representation(IR) compiled into multiple query languages, but it remains training‑based and lacks quality guarantees.}
Emerging Text2Cypher benchmarks~\cite{feng2024cypherbench} begin addressing Cypher generation, yet broad, fair cross‑language evaluation remains limited.
Our work fills this gap by introducing a constraint‑based IR refined through Chase \& Backchase within an \llm‑assisted agent pipeline, deploying Freebase graph on Amazon Neptune to activate mature SPARQL benchmarks for Cypher query generation and evaluation.

\section{Query Generation under Constraints}
\label{sec:problem}


\stitle{Knowledge Graphs and Queries. }
Let $G=(V,E)$ be a knowledge graph with entities $V$ and typed edges $E$ represented as triples $\langle s,p,o\rangle$.
A graph query $Q$ 
(e.g., Cypher) 
denotes a conjunctive pattern with filters; its answer set $Q(G)$ is the set of bindings that satisfy the pattern.
We assume a referenced \llm\ oracle that provides a \emph{reference answer} set $A$ and its grounding $G(A)\subseteq V$ for the given NL query $Q_n$.

\stitle{Constraints. }
We introduce a class of \emph{constraint tables} (\ctable) as a concise, language‑agnostic abstraction for graph queries, which contains: 
(i) \underline{\textit{typed triple patterns}} $\langle c_s,c_p,c_o\rangle$, where entries may be constants or variables; 
(ii) \underline{\textit{value constraints}} on literals (e.g., $x\,\texttt{op}\,c$, $\texttt{op}\in\{=,>,<,\ge,\le,\neq\}$); and 
(iii) \underline{\textit{structural constraints}}, such as joins, \texttt{OPTIONAL}/\texttt{UNION}, bounded paths, etc.
In practice, \ctable\ also captures \emph{implicit/pragmatic} conditions not explicit in the NL query (e.g., \texttt{Approved=True} in Fig.~\ref{fig:motivation}).
Each constraint $c$ can carry an uncertainty score $u\!\in\![0,1]$ reflecting specificity to guide search; constraints with no matches are pruned. 

\stitle{Quality Measures}. 
We consider \emph{relative} criteria w.r.t.\ the reference answers. 
We say a generated query $Q$ is 
(i) \underline{\textit{\llm-sound}}, iff. $Q(G)\subseteq G(A)$; 
(ii) \underline{\textit{\llm-complete}}, iff. $G(A)\subseteq Q(G)$.
(iii) \underline{\textit{Consistency}} iff. every binding in $Q(G)$ satisfies all $c\in\C$.
For monotone conjunctive queries, \textit{removing} constraints enlarges the answer set (hence may improve the answer completeness, yet by sacrificing soundness) while \textit{adding} constraints 
makes answers more specific 
(hence may improve soundness but risks sacrificing completeness).

\stitle{Problem Formulation. }
Given an NL query, graph $G$, a scored \ctable\ $\C$ extracted from the NL query, and a reference set $A$, 
we aim to generate a graph query $Q$ that
(i) is \textit{consistent} with $\C$;
(ii) jointly maximizes \emph{\llm-completeness} and \emph{\llm-soundness} (reported as EM/F1);
and (iii) is \textit{minimal} (no redundant constraints).

The foundation of our proposed unified solution is to 
search over subsets of $\C$ with a beam strategy, 
organized in three major phases: 
\textit{Chase} phase (relaxation for completeness), 
starting with a universal query that enforces all constraints, 
and \textit{Backchase} (tightening for soundness/minimality), 
terminating at minimal queries, such that adding one more constraint 
compromises soundness; and 
\textit{Rendering} the selected 
logical plan (conjunctive constraints) 
to the target query language. 

\section{\uqg: Unified Query Generation Framework}
\label{sec:method}

\begin{figure*}[tb!]
\centerline{\includegraphics[width=0.86\textwidth]{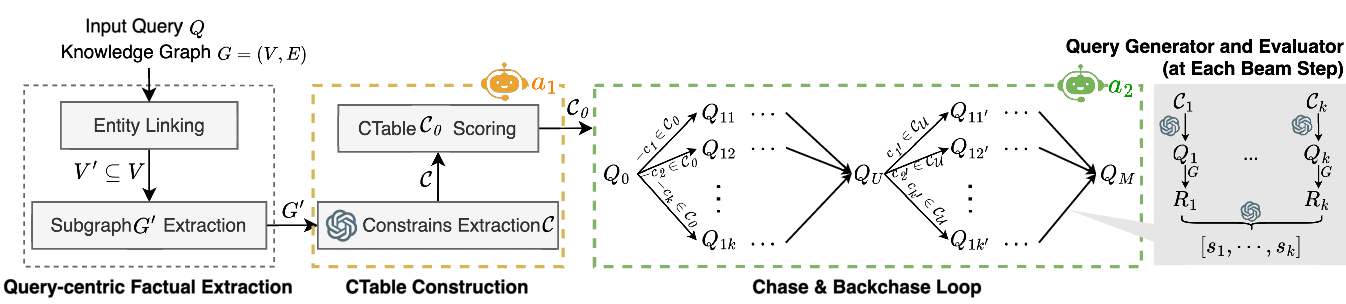}}
\centering
\vspace{-1ex}
\caption{\uqg Framework Overview}
\vspace{-3ex}
\label{fig:framework}
\end{figure*}

\uqg turns user intent into an executable query in three stages:
(i) query-centric fact extraction;
(ii) \ctable construction; and 
(iii) LLM-assisted 
Chase \& Backchase, 
followed by LLM‑prompted or deterministic rendering to the target query language.
Fig.~\ref{fig:framework} shows the end-to-end pipeline, with Agents~$a_1$ and $a_2$ orchestrating \ctable
construction and Chase \& Backchase, respectively.

\subsection{Query-centric Fact Extraction}
\label{sec:method:extract}

Given an input NL query $Q_n$ and a knowledge graph $G=(V, E)$, \uqg first performs entity linking to obtain
mentioned nodes $V' \subseteq V$ (using off-the-shelf
entity linkers~\cite{oliveira2021towards} 
or fact extractors~\cite{bekoulis2021review}).
For each $v\in V'$, it induces a local subgraph $G'(v)$ from the $k$-hop neighborhood (typically $k \leq 3$),
and unions them to form $G'=\bigcup_{v\in V'}G'(v)$. 
\revise{
The induced subgraph $G'$ serves as the factual context for \ctable construction; queries are executed on $G$ evaluated using $Q(G)$.}

\subsection{\ctable Construction}
\label{sec:method:ctable}

\stitle{Constraint Table Extraction}.
\agent $a_1$ constructs \ctable $\C$ by extracting relevant triples from $G$, focusing on the obtained subgraph $G'$.
$a_1$ samples and parameterizes triples from $G'$ based on their relevance to the input query $Q_n$. 
The triples are chosen for their ability to define meaningful relations. 
When hinted by $Q_n$, $a_1$ augments $\C$ with \emph{implicit/pragmatic} constraints
(e.g., \texttt{Approved=True} in Fig.~\ref{fig:motivation}).

To standardize processing, we normalize \ctable\ into a \emph{monotone core} that retains only positive atoms, constant filters, and bounded path existence; we rewrite
\texttt{OPTIONAL} as explicit branches and keep \texttt{UNION} as separate branches, while deferring non-monotone
constraints (negation/anti-join, at-most/exactly $k$, aggregates, \texttt{ORDER}/\texttt{LIMIT})
to post-validation.

\stitle{Constraint Table Scoring}. 
For each constraint $c_i \in \C$, \agent $a_1$ assigns an {\em uncertainty score} $u_i \in [0, 1]$ that reflects specificity: 
let $n_i$ be the number of matches of $c_i$ in $G$ and $m=\max_{c_j\in\C} n_j$, then
$u_i=\frac{n_i}{m}$.
Higher $u_i$ indicates a less specific (more permissive) constraint; constraints with $n_i=0$ are pruned.
These one-time scores guide beam priorities in the Chase and Backchase phase (Sec~\ref{sec:chase}).
We present the detailed procedure and cost analysis in Appendices~\ref{app:alg}-\ref{app:demo}~\cite{full}.

\subsection{Chase and Backchase Loop}
\label{sec:chase}

Given a scored constraint table $\C_0$ (normalized to the monotone core), and an oracle reference set $A$ for the NL query $Q_n$, \uqg runs a two‑phase generation (\textit{Chase} and  \textit{Backchase}~\cite{deutsch2006query}) to balance \textit{\llm-completeness} and  \textit{\llm‑soundness}, while preserving  \textit{minimality}.
\agent~$a_2$ comprises (i) a \gen that renders a candidate query from a constraint set; and (ii) an \eva that scores candidates against $A$, and applies post‑checks for any non‑monotone constraints.
To reduce cost and latency, all candidates in a beam step are evaluated in a \emph{single} batched \llm\ call.

\stitle{Query Chase Phase}.
Given a \ctable $\C_0$ constructed from the input natural language query $Q_n$, 
\agent $a_2$ performs the {\em Query Chase} phase to iteratively refine an initial graph query $Q_0$ derived from the full \ctable $\C_0$, using beam search to explore candidate queries and remove unnecessary constraints. 
The objective of \qc is to derive a {\em universal query} $Q_U$ that satisfies all necessary constraints in $\C_0$, ensuring the consistency 
and relative completeness. 
\qc coordinates 
beam search with a dynamically maintained 
search tree, where each node $v_Q$
refers to a candidate query $Q$ with 
the associated structure that tracks the following. 
(1) retrieved answer $r$, 
(2) a fraction of \ctable $Q.\C$ 
that contains the constraints 
yet to be enforced to refine the 
query $Q$, 
(3) a performance score $p$ 
to be estimated by \eva, 
and an overall score quality $s$ to be 
calculated for guiding the beam search. 

At each node $v_Q$, 
\qc spawns a set of children in the 
search tree, each referring to a new 
candidate query $v_Q'$ that is obtained by 
removing a constraint (a triple pattern) 
from $Q.\C$. The score $s'$ is 
then estimated for $Q'$ as  
\begin{equation}
s'(c, Q) = \alpha(1-c.u) + (1-\alpha)Q.p
\label{eq:score}
\end{equation}
Where $c.u$ is the uncertainty score of $c$, $Q.p$ is the performance score of the parent query, and $\alpha \in [0, 1]$ is a weighting factor to balance $c.u$ and $Q.p$. Indeed, 
the less constraints a query $Q$ 
poses, the more likely it ``covers'' 
ground truth, yet at a cost of introducing
more 
entities, and less precise answer. 

\begin{algorithm}[tb!]
\caption{\qc: completeness‑guided beam search}
\label{alg:chase}
\begin{algorithmic}[1]
\algtext*{EndFor}
\algtext*{EndIf}
\algtext*{EndWhile}
\algtext*{EndFunction}
\algtext*{EndProcedure}

\Statex \hspace{-4.5ex} \textbf{Input:} 
    NL Query $Q_n$; 
    \ctable $\C_0$; 
    \gen $\g$;
    \eva $\e$;
    Beam width $b$; 
    Threshold $\tau$;
    Graph $G$
\Statex \hspace{-4ex} \textbf{Output:} 
     Universal Query $Q_U$.
\vspace{0.6ex}

\State \textbf{set} $B := \varnothing$, $L := \varnothing$; 
\State \textbf{initialize} $s_0 := 0$, $B$.append(($\C_0$, $s_0$));
\vspace{0.6ex}

\While{$B \neq \emptyset$}
\For{$(\C, s) \in B$}
    \State $Q :=$ Gen$(\g, \C)$, $r := Q(G)$, $p:=0$;
    \State $L$.append$((\C, Q, r, p, s))$;
\EndFor
\State $L :=$ Eva($\e$, $L$, $Q_n$);

\State $l_U := \arg\max_{l \in L} l.p$, $Q_U := l_U.Q$; 
\If{$l_U.p \geq \tau$} break; \EndIf

\State $B := \{(l.\C \setminus \{c\}, s(c, l))  \mid l \in L, c \in l.\C\}$;

\If{$|B| \leq b$} continue;
\EndIf

\State $B := \{(\C, s) \in B \mid \text{top } b \text{ by } s\}$, $L := \varnothing$;
\EndWhile
\vspace{0.6ex}

\State \Return $Q_U$
\end{algorithmic}
\end{algorithm}

\eetitle{Outline}. 
The procedure executed by \agent $a_2$ is detailed in Alg.~\ref{alg:chase}.  
The beam $B$ encodes an initial query with all constraints in $\C_0$ as a conjunctive query, with the full \ctable $\C_0$ and a score $s_0=0$ (lns 1-2).
For each candidate $(\C, s) \in B$, \gen $\g$ generates a query $Q$, which is executed on $G$ to retrieve results $r$ (lns 4-5).
List $L$ is used to maintain all the 
relevant information for the candidate queries as a label $(\C, Q, r, p, s)$, where $p$ is initially a placeholder (ln 6).
After all candidates are processed, \eva $\e$ 
(1) verifies the \llm-relative completeness \wrt a reference answer set from an \llm oracle 
, as a hard constraint, and (2) assigns performance scores $p$ for each 
query (ln 7). 
If any $l.p$ exceeds the threshold $\tau$, the corresponding query $l.Q$ is returned as the universal query $Q_U$ (lns 8-9).
Otherwise, new candidates are spawned by removing one constraint at a time from $\C$ and calculating new scores $s$ for each descendant (ln 10). The candidates with top $b$ scores $s$ are selected for next beam step (lns 11-12).
Stop when a candidate reaches $p\ge\tau$ or the beam is exhausted; the best candidate so far is returned as the \emph{universal} plan $Q_U$.

\eetitle{\revise{Remarks}}. \revise{Since \ctable is normalized to the monotone core (Sec.~\ref{sec:problem}), dropping a constraint relaxes the candidate query and can enlarge its answer set.
Therefore, when a candidate fails the completeness check against the reference set $A$, removing constraints is a principled mechanism to recover missing answers.
We drop \textit{one} constraint at a time (Alg.~\ref{alg:chase}, ln~10) to enumerate \textit{minimal} relaxations and keep the branching factor controllable under a fixed beam budget.
Relaxation may introduce spurious answers and larger intermediate results; we mitigate this via uncertainty-aware beam scoring (Eq.~\ref{eq:score}) and subsequently tighten the universal query in \qbc.}

\stitle{Query Backchase Phase}.
In \qbc, \agent $a_2$ follows a bottom-up approach, starting with subqueries derived from the universal query $Q_U$, where each subquery corresponds to exactly one constraint from the \ctable $Q_U.\C$. These subqueries serve as the building blocks for constructing minimal queries that satisfy the relative soundness \wrt the answer set $A$.

\eetitle{Outline}. 
For each constraint $c_i \in Q_U.\C$, a subquery $Q_i$ is generated to cover only the single constraint $c_i$.
Similar to \qc (Alg.~\ref{alg:chase}), in {\em Backchase Phase}, \agent $a_2$ employs beam search to iteratively combine these subqueries. At each step, constraints are reintroduced from the subqueries, and candidate queries are evaluated based on their ability to produce valid results on the knowledge graph $G$.
\eva $\e$ is responsible for evaluating the soundness of the answer sets retrieved by candidate queries. The score $s$ is calculated (Eq.~\ref{eq:score}) to guide the beam search and select the most promising candidates.

\revise{
We present the full procedure in Appx.~\ref{app:alg}, and showcase a complete illustration in Appendix~\ref{app:demo}~\cite{full}.
}

\stitle{Quality guarantee}.
\revise{The \qc and \qbc together as an integrated 
framework eventually guarantee the consistency. 
The \qc phase (resp. \qbc) phase 
ensures relative completeness (resp. soundness) for refining 
the generated queries \wrt answer set $A$, provided either by the $NL$ queryer or generated by the \llm agent. These guarantees are ensured by guarding the correctness and completeness (resp. soundness) of the retrieved answers against $A$ at runtime. } 
Moreover, among all validated complete Cypher queries, the one that maximizes (resp. minimize) the number of 
constraints in $\C$ will be returned, ensured by the ``top-down''(resp. ``bottom-up'') strategy of \uqg.

\eetitle{\revise{Remarks}}. \revise{
The above quality guarantees are 
relative to a proper reference set. 
The latter can be obtained from 
crowd-sourced annotation, 
query-by-example, or querying 
high-quality master data. 
Fact-checking remains 
necessary yet out of scope 
of this work, and we leave it for 
future work. 
}

\subsection{Deployment Considerations}
\label{sec:method:deploy}
\vspace{-1ex}

\stitle{Runtime Environment \& Integration. }
Our framework runs on AWS EC2 \texttt{c5.4xlarge} instances (16 vCPU, 32\,GiB) behind an autoscaling group. 
We deployed Freebase (N-Triples, original support SPARQL only) on Amazon Neptune, supported by Neptune Analytics's MLM~\cite{schmidt2024opencypher} feature, enabling unified evaluation across both openCypher and SPARQL against the same graph using deployed endpoints without duplicating logic.
Agents are model-agnostic and implemented with LangChain. We invoke LLMs through Amazon Bedrock, which allows us to swap model families without code changes and provides an interface for inferring OpenAI models.

\stitle{Adaptation \& Portability. }
As \uqg is \emph{training-free}, adapting to a new KG or schema drift requires only updating synonym/hint lists; the agents will discover the updated knowledge bases at runtime as described.
The query rendering layer is pluggable: the same plan can be emitted as Cypher or SPARQL via a mapping set. 
When any constraint lacks a direct mapping, we fall back to prompt-based code generation and gate acceptance via answer-set validation, which keeps cross-language evaluation feasible.

\stitle{Cost Control \& Guardrails. }
To keep per-query cost predictable and tail latency bounded, \uqg (i) batches \eva into one LLM call per beam step for fairness and efficiency scoring; 
(ii) parallelizes candidate generation and execution within each beam step;  
(iii) bounds beam width/depth and adapts them using early completeness signals;
(iv) short-circuits empty/degenerate plans; and
(v) caches entity links, $k$-hop subgraphs, and reference answer sets.
We also enable beam parameters auto-tuning to fit the cost budget.

\stitle{Onboarding Artifacts. }
We provide a one-step deployment recipe (IaC) and a well-curated graph snapshot to load Freebase into Amazon Neptune, build indices, and provision endpoints;
a harness to run common Freebase-based KGQA benchmarks 
in \emph{dual-language} mode (Cypher/SPARQL);
and \emph{gold Cypher pairs} aligned with popular SPARQL benchmarks to reduce cold-start for LPG/Cypher-based KGQA.

\noindent\revise{
\stitle{Summary. } These considerations translate the industrial pain points in Sec.~\ref{sec:intro} into concrete guardrails:
(i) \textit{integration and portability} via Neptune dual-language endpoints and model-agnostic agents;
(ii) \textit{robustness under schema mismatch/drift and LLM mis-grounding} via query-centric extraction, a pluggable rendering layer, and CaB refinement;
and (iii) \textit{predictable latency/cost under SLAs} via batched evaluation, parallel execution, bounded beam width/depth with early stopping, short-circuiting degenerate plans, and caching.
The provided IaC recipe and benchmark harness further reduce cold-start and maintenance overhead in practice.}

\section{Experiment Study}
\label{sec:exp}

\subsection{Experiment Settings}
\label{sec:exp:setting}

\stitle{Datasets}. 
To mirror the real-world enterprise KGQA scenarios our Amazon team targets - querying production‑scale graphs, we select the \free as knowledge base ($2.9$B triples, $116$M entities~\cite{zheng2018question}) and adopt three well-established benchmarks that span everyday, multi‑hop, and schema‑diverse questions:
\revise{
(i) \underline{\gq}~\cite{su2016graphquestions}, a characteristic-rich testbed with 2,381 training and 2,395 test questions
(train:test $\approx$ 0.99);
(ii) \underline{\grail}~\cite{gu2021beyond}, covering three generalization settings (i.i.d., compositional, zero-shot),
with 44,337 training questions and a dev split of 6,763 questions (train:test $\approx$ 6.56); and 
(iii) \underline{\webqsp}~\cite{yih2016value}, with 3,098 training and 1,639 test questions (train:test $\approx$ 1.89).}
As \uqg is training-free, we use only the test sets, which consist of $2395$, $6763$, and $1639$ questions, respectively.

\stitle{Methods}.
(i)
\underline{\uqg}:
our unified pipeline, where ``\textit{Universal}'' refers to the results from the universal query, “\textit{Minimal}” refers to the minimal query, and ``\textit{Optimal}'' selects the better of the two per question, evaluated with GPT-4o;  
(ii) 
\underline{\blone}: a prompt‑only variant using the same prompt as \gen in Agent \(a_2\), evaluated with GPT-4o;
(iii) \underline{\arcane}~\cite{gu2022arcaneqa}: a generation model combining program induction and contextual encoding, SPARQL only; and
(iv) 
\underline{\pangu}~\cite{gu2023don}: a state‑of‑the‑art framework with a symbolic agent for exploration and a neural agent for evaluation, models fine-tuned with specific benchmarks and schema, SPARQL only. 
To ensure a fair comparison and isolate entity-linking effects, we adopt the same entity-linking results as \pangu.

\stitle{Evaluation Metrics}
We evaluated the quality of generated queries with the following metrics, against ground truth answers provided in benchmarks:
(i) \underline{\exact},  the proportion of queries that return exactly the same answer as the ground truth; 
(ii) \underline{Precision(P)}, \underline{Recall(R)}, and \underline{\f}: \textit{Average} scores for all queries in terms of their completeness and soundness of the returned answers; and (iii) query‑generation latency, measured in per‑query time, reported as median (P50) and tail (P95) metrics. P50 denotes the median latency, and P95 is the value below which 95$\%$ of the queries complete.

\noindent\revise{
\stitle{Evaluation protocol.}
All methods query the same Neptune-hosted Freebase snapshot.
\arcane and \pangu are SPARQL-only baselines, while \uqg is evaluated under both SPARQL and openCypher renderings.
We compute EM/P/R/F1 on answer sets using the same evaluation script for all settings.
To reduce confounding effects from entity linking, UniQGen reuses the off-the-shelf entity-linking outputs provided by \pangu.
\arcane and \pangu are schema-aware methods that are tuned with KG schema/ontology, whereas UniQGen and the \blone are schemeless methods. All reported scores are evaluated against the benchmark ground‑truth answers rather than $A$.}


\begin{table*}[tb!]
\centering
\begin{adjustbox}{max width=\textwidth}
\begin{tabular}{c|cccc|cccc|cccc|cccc}
\toprule
\multirow{2}{*}{\textbf{Method}} 
& \multicolumn{4}{c|}{\textbf{I.I.D.}} 
& \multicolumn{4}{c|}{\textbf{Compositional}} 
& \multicolumn{4}{c|}{\textbf{Zero-shot}} 
& \multicolumn{4}{c}{\textbf{Overall}} \\
\cmidrule(lr){2-5}\cmidrule(lr){6-9}\cmidrule(lr){10-13}\cmidrule(lr){14-17}
& P & R & F1 & EM & P & R & F1 & EM & P & R & F1 & EM & P & R & F1 & EM \\
\midrule
\blone\ 
& 0.291 & 0.293 & 0.291 & 0.276 
& 0.300 & 0.312 & 0.303 & 0.277 
& 0.313 & 0.313 & 0.313 & 0.310 
& 0.305 & 0.308 & 0.306 & 0.295 \\
\arcane
& 0.887 & 0.896 & 0.887 & 0.869 
& 0.744 & 0.763 & 0.739 & 0.690 
& 0.728 & 0.741 & 0.727 & 0.706 
& 0.769 & 0.782 & 0.767 & 0.741 \\
\pangu\ 
& 0.892 & 0.900 & 0.891 & 0.869 
& \underline{0.790} & \underline{0.803} & \textbf{0.787} & \underline{0.738} 
& 0.827 & 0.835 & 0.826 & 0.812 
& 0.834 & 0.843 & 0.833 & 0.809 \\

\lightmidrule

\uqg-\uni\
& \underline{0.908} & \underline{0.909} & \underline{0.903} & \underline{0.886}
& 0.763 & 0.779 & 0.747 & 0.702
& \underline{0.896} & \underline{0.895} & \underline{0.893} & \underline{0.885}
& \underline{0.869} & \underline{0.872} & \underline{0.863} & \underline{0.844} \\
\uqg-\mini\
& 0.903 & 0.905 & 0.898 & 0.883
& 0.740 & 0.759 & 0.727 & 0.682
& 0.885 & 0.886 & 0.882 & 0.874
& 0.857 & 0.862 & 0.851 & 0.833 \\
\uqg-\all\
& \textbf{0.914} & \textbf{0.912} & \textbf{0.908} & \textbf{0.893}
& \textbf{0.792} & \textbf{0.807} & \underline{0.779} & \textbf{0.740}
& \textbf{0.900} & \textbf{0.899} & \textbf{0.898} & \textbf{0.890}
& \textbf{0.879} & \textbf{0.881} & \textbf{0.874} & \textbf{0.857} \\
\bottomrule
\end{tabular}
\end{adjustbox}
\caption{\small{GrailQA results by split; compare \uqg in Cypher with SPARQL-only methods; best per column in \textbf{bold}, 2nd best \underline{underlined}.}}
\label{tab:grailqa}
\vspace{-1ex}
\end{table*}

\begin{figure*}[tb]
\centering
\begin{minipage}[t]{0.63\textwidth}
\vspace{-32.7ex}
  \begin{adjustbox}{max width=\textwidth}
  \begin{tabular}{c|c|cccc|cccc}
\toprule
\multirow{2}{*}{\makecell{\textbf{Query}\\\textbf{Language}}} & \multirow{2}{*}{\textbf{Method}}
& \multicolumn{4}{c|}{\textbf{GraphQ}} 
& \multicolumn{4}{c}{\textbf{WebQSP}} \\
\cmidrule(lr){3-6}\cmidrule(lr){7-10}
& & P & R & F1 & EM & P & R & F1 & EM \\
\midrule
\multirow{5}{*}{\textsc{SPARQL}}
& \arcane 
& 0.346 & 0.400 & 0.343 & 0.320 
& 0.753 & 0.781 & 0.748 & 0.706 \\
& \pangu\ 
& 0.614 & 0.655 & 0.610 & 0.584 
& \underline{0.792} & \underline{0.821} & \underline{0.793} & \underline{0.758} \\
& \uqg-\uni
& 0.750 & 0.789 & 0.756 & 0.740 
& 0.785 & \textbf{0.937} & \textbf{0.794} & 0.648 \\
& \uqg-\mini
& 0.753 & 0.792 & 0.759 & 0.744
& 0.785 & \textbf{0.937} & \textbf{0.794} & 0.647 \\
& \uqg-\all 
& 0.763 & 0.805 & 0.769 & \underline{0.752}
& 0.785 & \textbf{0.937} & \textbf{0.794} & 0.648 \\
\lightmidrule
\multirow{4}{*}{\textsc{Cypher}}
& \blone 
& 0.221 & 0.225 & 0.223 & 0.214 
& 0.290 & 0.270 & 0.270 & 0.260 \\
& \uqg-\uni
& \underline{0.778} & \underline{0.810} & \underline{0.773} & 0.740
& -- & -- & -- & -- \\
& \uqg-\mini
& 0.747 & 0.776 & 0.744 & 0.714
& -- & -- & -- & -- \\
& \uqg-\all
& \textbf{0.809} & \textbf{0.835} & \textbf{0.803} & \textbf{0.775}
& \textbf{0.793} & 0.818 & 0.788 & \textbf{0.781} \\
\bottomrule
\end{tabular}
  \end{adjustbox}
  \captionof{table}{\small{\gq \& \webqsp results across SPARQL and Cypher; all queried over the same Neptune-managed endpoint; best per column in \textbf{bold}, second best \underline{underlined}.}}
  \label{tab:graphq-webqsp}
\end{minipage}%
\hfill
\begin{minipage}[b]{0.347\textwidth}
  \centering
  \includegraphics[width=\linewidth]{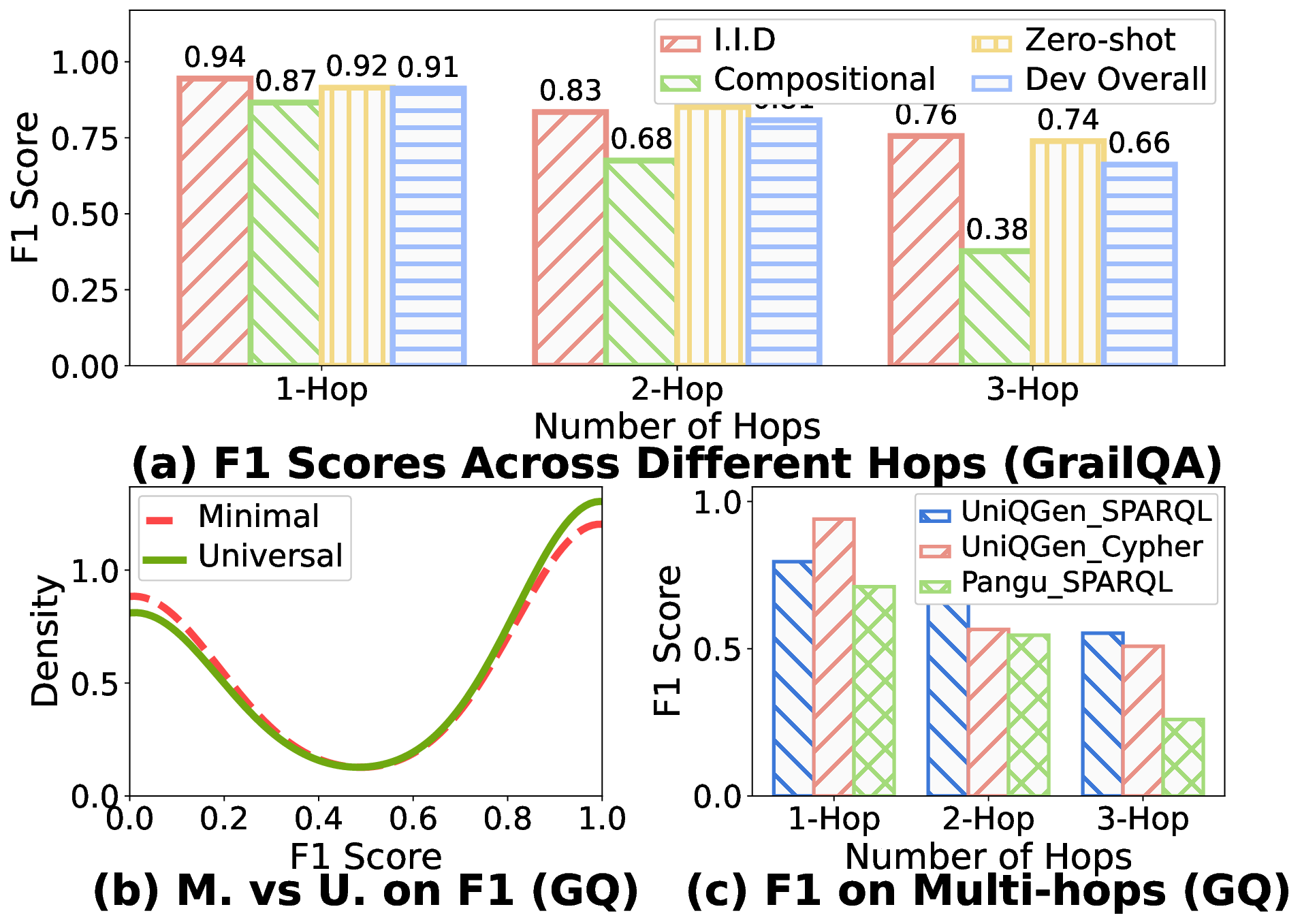}
  \captionof{figure}{\revise{Effectiveness \& Query Quality}}
  \label{fig:performance}
\end{minipage}
\vspace{-2ex}
\end{figure*}

\vspace{-.8ex}
\subsection{Results and Analysis}
\label{sec:exp:results}
\vspace{-1ex}

\stitle{Effectiveness \& Query Quality}. 
We summarize query quality results in Table~\ref{tab:grailqa} (GrailQA) and Table~\ref{tab:graphq-webqsp} (GraphQ/WebQSP). \uqg consistently surpasses baselines across all datasets and query types, notably on \gq, highlighting its effective constraint exploration without requiring costly model training or fine-tuning. This efficiency and adaptability showcase its industrial scalability for diverse query scenarios.



\eetitle{Accuracy Variation Across Datasets.}
\revise{
\uqg sees the largest gains on \gq, the most low-resource benchmark among the three, with the smallest train:test ratio ($\approx$ 0.99). This setting constrains training-based baselines more severely, while \uqg remains training-free and thus generalizes better. We further diagnose \gq by query complexity (Fig.~\ref{fig:performance}(c)): while all methods degrade as hops increase, \uqg degrades more gracefully than \pangu, and the gap widens on multi-hop questions, indicating that constraint-guided search and execution-based validation particularly help compositional queries under limited supervision.
The trend also remains consistent across both renderings, mitigating concerns that the gain is renderer-specific.}
In contrast, \webqsp is more diverse and amenable to in-context learning, shows \uqg achieving results comparable to or slightly below \pangu. On \grail, \uqg consistently leads on I.I.D. and Zero-shot subsets but slightly trails on Compositional due to complex reasoning, which benefits more from few-shot adaptation. 
Interestingly, the winning renderer flips by dataset: Cypher leads on \gq while SPARQL leads on \webqsp. This aligns with schema bias: \gq’s patterns map cleanly to Cypher’s labeled-node/edge model and bounded relationship lengths, while \webqsp frequently touches Freebase’s RDF specifics (mediator/CVT nodes, directional predicates, typed literals), which are most naturally expressed with SPARQL triples and property paths.


\eetitle{Complexity of Input Query Classes (Multi-Hop).} 
Fig.~\ref{fig:performance}(a) shows declining F1 scores with increasing hops, especially noticeable in the compositional subset ($56\%$ drop from 1-hop to 3-hop). Multi-hop reasoning complexity inherently poses challenges, as it requires integrating information across multiple relationships.
Despite \uqg being training-free, it maintains competitive performance in multi-hop scenarios in most cases, confirming the practicality of constraint-guided planning for complex patterns in real-world applications.

\eetitle{F1 Score Distribution: Insights. } 
Fig.~\ref{fig:performance}(b) shows a bimodal distribution, with peaks near $0$ and $1$, which reflects the effect of Chase \& Backchase Loop: when \ctable fails to capture critical constraints, early pruning yields low scores; when aligned, following Chase \& Backchase refinement effectively boosts soundness and completeness, yielding high scores. 
Notably, Universal queries tend to outperform Minimal ones, validating our Chase-first design for completeness.


\begin{figure}[t!]
  \centering
  \includegraphics[width=0.49\textwidth]{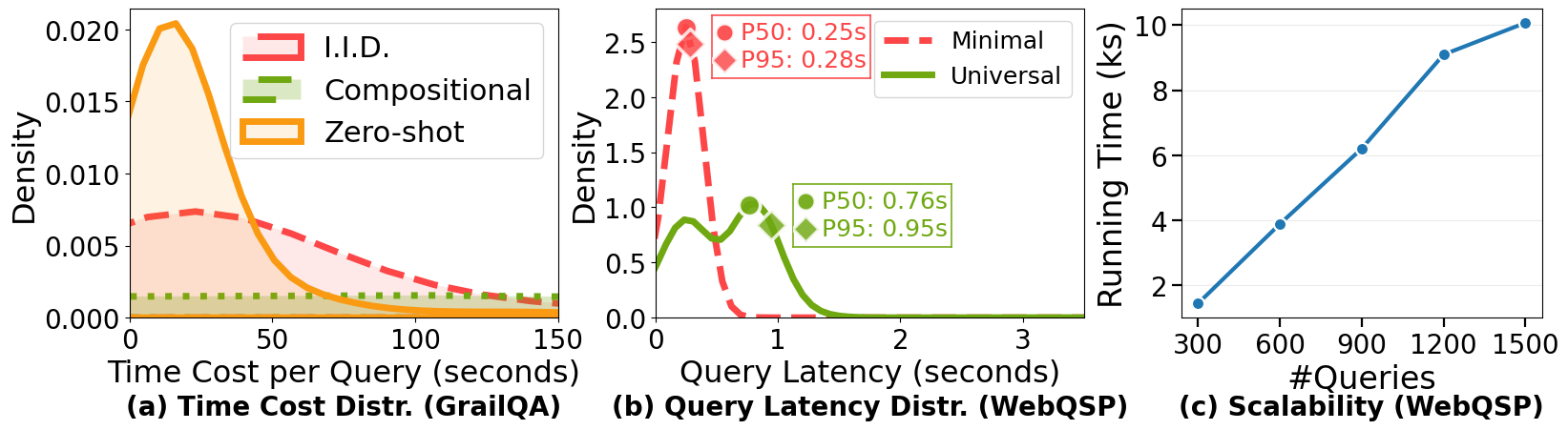}
  \vspace{-3ex}
  \caption{Efficiency Analysis}
  \label{fig:overhead}
  \vspace{-4ex}
\end{figure}

\stitle{Efficiency. }
Unlike training-based systems that typically require \textit{multi-days} fine-tuning on a Freebase-scale graph~\cite{pangu}, \uqg has \textit{no} training cost; the deployment overhead is purely on runtime planning and 
query execution.
As shown in Fig~\ref{fig:overhead}(a), \revise{
it takes in total on average around $40$ seconds per query generation, outperforming 
most baselines. 
}
This suggests that \uqg is feasible even for intricate multi-hop queries. Additionally, queries from the Compositional subset require longer processing times, 
which is consistent with the impact of more complex queries in that set.

\revise{
We use \webqsp as a public proxy for user-facing industrial queries and evaluate \uqg under production-style constraints on our Neptune-hosted Freebase endpoint (timeout=60s, beam width/max\_depth=5/5, caching=ON).
To ensure label consistency with our execution setting, we re-execute the dataset-provided gold SPARQL queries on the same endpoint and use the resulting answer sets as business labels, with a successful re-execution rate of $89.69\%$.
On this verified workload, \uqg achieves EM/F1 of 0.78/0.79; Fig.~\ref{fig:overhead}(b) reports \emph{KG execution latency} of the generated plans, where the Minimal plan achieves P50/P95 of 0.25s/0.28s and the Universal plan remains under 1s at P95.
We further report per-query \emph{resource cost}: token usage (1570.43 input/1823.29 output), per-query 2 LLM calls, and average 17 KG executions, showing a practical quality, latency, and cost trade-off.
We also add a scalability analysis under query load on WebQSP (Fig.~\ref{fig:overhead}(c)), showing that \uqg’s end-to-end generation latency remains stable up to at least 1500 concurrent requests, indicating predictable performance under production-style constraints.
}

\revise{We also report an ablation study to justify 
critical design choices and more real-world 
use case. Due to limited space, we 
present our findings in the Appendix.}

\section{Conclusions}
\label{sec:conclusion}
\vspace{-1ex}

We have introduced \cc, a unified solution 
to generate graph queries from natural language 
inputs for KGQA tasks. We have 
highlighted its principled deployment-friendly 
framework based on a variant of Chase \& Backchase 
process that optimizes the satisfaction of constraints 
obtained by \llms, quantified by relative soundness 
and completeness measures over reference answer sets. 
Our experimental study 
has verified its ability to generate high-quality 
Cypher and SPARQL queries, and its scalability 
over large datasets and queryload. 



\balance
\bibliographystyle{IEEEtran}
\bibliography{paper}

\setcounter{section}{0}
\titleformat{\section}[block]
{\normalfont\large\bfseries}{Appendix \Roman{section}:}{0.5em}{}
\newpage
\onecolumn

\section{\one{Glossary of Key Notations}}
\label{app:glossary}

* The examples related to the Olympic Games are based on the context provided in Example~\ref{exa-cab}.

\begin{description}

\item[$Q_n$] – Natural-language query/question (input).

\item[$Q$] 
-- A structured Cypher query generated from a natural language query. \\
\textit{Example:} \texttt{MATCH (c:City)-[:hosted]->(e:OlympicGames)}

\item[$G$] 
-- Knowledge graph containing entities (nodes) and relations (edges).

\item[$Q(G)$] 
-- Answer set obtained by executing query $Q$ on $KG$.
\revise{
\item[$V'$] 
-- Entity-linked node set extracted from $Q_n$, $V' \subseteq V$;}
\revise{
\item[$G'$]
-- Query-induced context subgraph for constraint extraction, $G'=\bigcup_{v\in V'}G'(v)$;}

\item[$A$]
-- The reference answer set provided by an oracle LLM in response to a natural language query. \\
\textit{Example:} For "Olympics cities in the USA," $A$={``Atlanta'',``St.Louis'',``LosAngeles''}.

\item[$G(A)$]
-- All $KG$ entities corresponding directly to answers in the reference set $A$.


\item[$\C$ (\ctable)]
-- A structured representation (Constraint Table) of all constraints, capturing triple patterns extracted from $KG$ to support accurate query generation. \\
\textit{Example:} Entries: \textit{(City, hosted, OlympicGames)}, \textit{(OlympicGames, year, >1896)}.

\item[$c \in \C$]
-- An individual constraint within the \ctable $\C$, defined as a triple pattern or condition. \\
\textit{Example:} \textit{(City, hosted, OlympicGames)} (single constraint).

\item[$u$]
-- Constraint uncertainty score (0–1), quantifying the specificity of each constraint. Lower $u$ means higher specificity. \\
\textit{Example:}
Constraint \textit{(year>1900)} have a higher uncertainty score than \textit{(year>2000)}, indicating broader applicability.

\item[$p$]
-- Performance score assigned by the Evaluator to a query, measuring how well its result matches the oracle answer set $A$. \\
\textit{Example:}
If query results fully match Oracle answers 
$A$, then $p \approx 1.0$

\item[$s$]
-- Overall score for candidate queries used to rank queries during beam search, combining constraint uncertainty $u$ and query performance $p$.

\item[$Q_U$ (Universal Query)]
-- Query derived in the Chase phase, ensuring completeness by initially including all constraints from $\C$. \\
\textit{Example:} A universal query retrieves all cities hosting Olympics (both Summer and Winter), maximizing completeness.

\item[$Q_M$ (Minimal Query)]
-- Query generated in the Backchase phase by minimizing constraints from $Q_U$, ensuring soundness. \\
\textit{Example:} 
A minimal query includes only essential constraints, e.g., filters specifically to "Summer Olympics.".

\item[$b$ (Beam Width)]
-- Number of candidate queries retained at each iteration (beam step) during Chase $\&$ Backchase processes. \\
\textit{Example:} With $b = 5$, only the top $5$ candidates by score $s$ proceed to the next iteration.

\end{description}

\section{Algorithms}
\label{app:alg}

\subsection{\three{\ctable Construction}}

\begin{algorithm}[H]
\caption{CTable Construction}
\label{alg:ctable} 
\begin{algorithmic}[1]
\algtext*{EndFor} 
\algtext*{EndIf} 
\algtext*{EndWhile} 
\algtext*{EndFunction} 
\algtext*{EndProcedure}

\Statex \hspace{-4.5ex} \textbf{Input:}  Natural language query $Q_n$, Knowledge Graph $G$, hop size $k$, Match Cap $\tau$, \agent $a_1$
\Statex \hspace{-4ex} \textbf{Output:} Scored Constraint Table $\mathcal{C}_0$

\State $E \gets$ EntityLinking($Q_n$)
\Comment{Link entities in the NL query}

\State $G' \gets$ Subgraph($G$, $E$, $k$)
\Comment{Extract $k$-hop subgraph}


\State \textbf{Initialize} \ctable $\mathcal{C} := \emptyset$

\For{each triple $t = \langle s, p, o \rangle$ in $G'$}
    \If{$t$ is relevant to the $Q_n$}
        \State $\mathcal{C} := \mathcal{C} \cup \{t\}$
    \EndIf
\EndFor

\State Extend $\C$ by using $a_1$ to inject additional constraints (e.g., filters, type conditions)

\For{each constraint $c_i \in \mathcal{C}$}
    \State $n_i \gets$ MatchCount($G$, $c_i$) \Comment{Count supports in $KG$}
\EndFor

\State \textbf{Let} $m := \min(\max_i n_i, \tau)$ 

\For{each $c_i \in \mathcal{C}$}
   \State $u_i := \frac{n_i}{m}$ \Comment{Compute uncertainty score}
\EndFor

\State $\mathcal{C}_0 := \{(c_i, u_i) \mid u_i > 0\}$
\Comment{Prune constraints with zero matches}

\State \Return $\mathcal{C}_0$

\end{algorithmic}
\end{algorithm}

Algorithm~\ref{alg:ctable} describes the procedure used by \agent $a_1$ to construct a scored constraint table (\ctable) from a natural language query $Q_n$. 
Initially, EntityLinking function identifies entities $E$ explicitly mentioned in $Q_n$. Then, a relevant subgraph $G'$ is extracted from the knowledge graph $G$ by retrieving nodes and edges within $k$-hop neighborhoods of these linked entities. 
Relevant triples within $G'$ that align with the query intent are collected to form the initial \ctable, which \agent $a_1$ subsequently enriches by injecting inferred constraints, such as additional type restrictions and filtering conditions. 
Each constraint $c_i$ in the \ctable is evaluated by counting its support $n_i$, the number of matching triples in $G$. 
An uncertainty score $u_i$ is then computed as the normalized support count relative to the maximum support count (capped by $\tau$). 
Constraints without any matches are pruned, resulting in the final scored constraint table $\C_0$.

\subsection{Query Backchase}

We present more details of 
Query Backchase, as shown 
in Algorithm~\ref{alg:backchase}. Similar as its Chase 
counterpart, Backchase works with 
(1) a query Generator $\gamma$, to explore 
sub-queries of the input universal query $Q_U$ following a bottom-up 
manner (by``merging'' from small, 
single-edge patterns into larger 
sub-queries); and (2) 
an Evaluator $\epsilon$ that consults 
an \llm to identify answers by 
evaluating 
an input sub-query $Q$ generated from $\gamma$. The algorithm follows a 
beam search process, 
starting with the universal query $Q_U$, 
and initializes a set of 
sub-queries by decomposing 
$Q_U$ into triple patterns 
$c$ (line~2). It coordinates a 
the backchase then by generating 
and verifying a re-combination 
of these sub-queries into 
larger counterparts (lines 4-8), 
by ``adding back'' the 
constraints as part of the 
subqueries from the extracted 
constraint table \ctable. 
The Evaluator (line 8), 
during the beam search 
based exploration of 
sub-queries, ensures the 
hard constraints by verifying  
if the current subquery $Q_n$ 
remains to be \llm-relative 
completeness, as a hard constraint. 
As is, the algorithm ensures 
an invariant that 
it only explores sub-queries that 
are guaranteed to be consistent and 
\llm-relative complete. 

\noindent\revise{
\stitle{Consistency Guarantee}. 
We clarify that the above 
process ensures a consistency guarantee 
for the output query $Q$ and its 
associated CTable $\C$. This is 
not a strong consistency guarantee 
which ensures $Q$ is always consistent with 
the input constraint table $\C_0$, yet a 
pragmatic guarantee to have reasonable 
query output, especially when 
$\C_0$ may be ambiguous or 
inconsistent -- in the latter case, no 
query with answers can simultaneously 
satisfy all the constraints at the same time.} 

\revise{
We consider 
a pair $(Q, Q.\C)$, initialized as 
$(Q_U, Q_U.\C_0)$. It suffices to show 
that the process has the invariant 
that for any output $(\C, Q)$ with 
updates that change $\C_0$ to $\C$ and 
$Q_U$ to $Q$, $Q$ has answer that 
satisfy all the constraints in $Q.\C$. 
This can be shown by an inductive 
analysis, for a reasoning path 
as a sequence of updates in Chase and 
BackChase to the pair $(Q, Q.\C)$. 
(1) The Query Chase Phase 
initializes CTable $\C_0$ with all 
constraints and a universal query $Q_U$ 
that contain all constraints in $\C_0$. 
If it outputs $(Q_U, Q_U.\C_0)$ without 
any Chase or BackChase step, then 
either $\C_0$ is $\empty$ (as $B$=$\empty$, line 3 
of Algorithm 1), or $Q_U$ contains a set of 
query nodes without any edges \ie 
no triple constraints can be enforced with an answer that satisfies all $\C_0$ -- a case that $\C_0$ 
is inconsistent. In either case, 
$Q_U$ is trivially consistent with $\C_0$. 
(2) Assume at step $i$ of the search, 
Chase derives $(Q, Q.\C)$ where $Q$ is consistent 
with $Q.\C$. At any spawning step $i+1$ in QChase, $Q$ 
is updated to $Q'$ by removing a triple pattern $c$ 
from $\C$. Consider the pair $(Q', Q'.\C)$, 
we have $Q(G)\subseteq Q'(G)$, for a fixed set 
of node variables. There are two cases: 
(a) the new answers in $Q'(G)\setminus Q(G)$ 
also satisfy $Q'\C$. In this case, $(Q', Q'.\C)$ 
ensures consistency guarantee. 
(b) at least one answer in $Q'(G)\setminus Q(G)$ 
does not satisfy $Q'.\C$ due to the removal of 
the triple pattern $c$. Then there must exist a 
step in QBackChase, posed on the subquery $Q_i$ 
that either is itself $c$, or a subquery $Q''$ of $Q'$ such 
that $Q''$ is derived by adding back $c$, with verified answer $Q''(G)$ that satisfy $Q''.\C''$. In both cases, 
it eventually outputs $(Q'', Q''.\C'')$ 
that satisfy the consistency guarantee. 
This proves the consistency 
guarantee. 
}

\noindent\revise{
\stitle{Relative Completeness Guarantee (QChase)}. 
We provide a similar inductive proof 
to show the completeness guarantee ensured by 
QChase phase, following our construction in the 
consistency analysis. 
(1) Given the reference set $A$, 
if QChase output $(Q_U, \emptyset)$, 
\ie $\C$ has no triple pattern to be 
removed, and $Q_U$ is a set of query nodes 
without edge constraint -- hence include all 
the nodes in $G$ with a simple type match only, hence $A\subseteq Q_U(G)$. 
(2) Assume at step $i$, $(Q, Q.\C)$ 
ensures relative completeness, \ie $A\subseteq Q(G)$. Then, at any follow-up step that 
spawns a new query from $Q$ and yields 
$Q'$ by removing a triple constraint, we have $A\subseteq Q(G) \subseteq Q'(G)$ because the removal of a triple constraint 
does not reduce the matched answers. Hence 
the relative completeness is ensured 
at any step of QChase. Note that 
QChase alone does not necessarily satisfy 
relative soundness, as relaxing constraints 
may introduce new answers not in $A$. This 
is coped with QBackchase. 
}

\noindent\revise{
\stitle{Relative Soundness Guarantee (QBackChase)}. 
Following our construction in the 
consistency and relative completeness analysis, 
consider the initialization of 
QBackChase that starts with a set of pairs 
$(Q_i, C)$, where each $Q_i$ is a single 
edge query of the input query $Q_U$ from QChase. 
As $(Q_U, Q_U.\C)$ ensures relative completeness, given the proof of relative completeness 
guarantee of QChase, we have 
$A\subseteq Q_U(G)$. Clearly for 
each $Q_i$ as a subquery of $Q_U$, 
$A\subseteq Q_U(G)$. 
The QBackChase proceeds by combining 
subqueries together and outputs 
once it verifies that 
$Q'$ = $\bigcup Q_j$, where 
$Q'(G)\subseteq A\subseteq Q_j(G)$, 
given the corresponding constraint $Q'.C'$ 
as the {\em conjunction}, \ie 
$Q'.\C$ = $\bigwedge Q_j.\C_j$. 
Hence, for every output 
$(Q', Q'.\C')$, the relative soundness 
holds. This completes the proof of 
relative soundness guarantee. 
}

\stitle{Minimality guarantee}. We advocate a desirable property as follows. 
Given a NL query $Q_0$ with 
a reference answer $A$ from an \llm, and 
a set of constraints encoded in a 
\ctable $\C$, 
a Cypher query $Q$ is a {\em minimal 
complete rewriting} of $Q_0$, 
if (1) $Q$ is \llm-relevant 
complete that is consistent 
with $\C$, and (2) for any 
sub-query obtained by removing 
a constraint from $Q_0$ in $\C$, 
$\C$ is no longer \llm-relevant 
complete. The minimality 
property ensures that 
the generated query 
contains a set of rich and accurate 
search conditions (constraints) 
that can accurately express 
the original NL queries, 
as close as possible. 

We verify that  
\qbc ensures to produce 
minimal complete rewriting 
of an input NL query $Q_0$. 
To see this, observe the following. 
(1) The sub-queries, as a conjunct 
of single constraint patterns 
(single-edge queries), 
are either discarded in 
the beam search due to that 
they return 
incomplete or empty answers, 
or are kept to be more accurate (selective) 
by enriching with more relevant 
constraints, which remains 
consistent, and \llm-relevant complete.
(2) For any children spawned from  
any returned query $Q$, 
\qbc ensures to discard them 
due to the violation of \llm-relevant 
complete. 

For time cost, Backchase incurs 
the same cost as its Chase counterpart, 
in $O(L(B+ |\C|\log|\C|))$ time, where $B$ is the 
beam size, and $L$ refers to the 
number of levels (the depth) the beam search 
goes. We remark that $|\C|$ is often a 
small constant, as it only contains relevant constraints derived from NL query $Q_0$ and 
relevant triples in KG, at query-time; hence in practice the time cost is comparable to $O(BL)$. Note that \qbc is also training-free, hence \cc does not incur additioinal learning 
overhead. 

\begin{algorithm}[tb!] 
\caption{Query Backchase (\qbc)} 
\label{alg:backchase} 
\begin{algorithmic}[1] 
\algtext*{EndFor} 
\algtext*{EndIf} 
\algtext*{EndWhile} 
\algtext*{EndFunction} 
\algtext*{EndProcedure}

\Statex \hspace{-4.5ex} \textbf{Input:} 
Universal Query $Q_U$; 
NL Query $Q_n$; 
\gen $\g$;
\eva $\e$; 
Beam width $b$; 
Threshold $\tau$; 
Graph $G$;

\Statex \hspace{-4ex} \textbf{Output:} Minimal Query $Q_M$. 
\vspace{0.6ex}

\State \textbf{set} $B := \varnothing$, $L := \varnothing$; 
\For {\textbf{each} $c_i \in Q_U.\C$}
\State $\C_i := \{c_i\}$, $B$.append($(\C_i, 0)$);
\EndFor
\vspace{0.6ex}

\While{$B \neq \emptyset$}
\For{$(\C, s) \in B$}
    \State $Q :=$ Gen$(\g, \C)$, $r := Q(G)$, $p:=0$;
    \State $L$.append$((\C, Q, r, p, s))$;
\EndFor
\State $L :=$ Eva($\e$, $L$, $Q_n$);
\Comment{calcu. $l.p$, $\forall l \in L$}

\State $l_M := \arg\max_{l \in L} l.p$, $Q_M := l_M.Q$; 
\If{$l_M.p \geq \tau$} break; \EndIf

\State $B :=$ \{{\scriptsize $(l.\C \cup \{c\}, s(c, l)) \mid l \in L, c \in (Q_U.\C \setminus l.\C)$}\}

\If{$|B| \leq b$} continue;
\EndIf

\State $B := \{(\C, s) \in B \mid \text{top } b \text{ by } s\}$, $L := \varnothing$;
\EndWhile
\vspace{0.6ex}

\State \Return $Q_M$ 
\end{algorithmic} 
\end{algorithm}

\revise{
\subsection{Design intuition (line~10 in Alg.~\ref{alg:chase})}
\label{app:alg:design}
}

\revise{
\textit{-- Why are new candidates spawned by removing one constraint at a time in \qc?}
}

\revise{
Alg.~\ref{alg:chase} (\qc) performs a top-down beam search over subsets of constraints in the scored
\ctable\ $C_0$ (normalized to the monotone core).
When a candidate query $Q$ fails the completeness check against the oracle reference set $A$,
\qc relaxes $Q$ by dropping constraints.
Under the monotone-core normalization, removing a constraint can only enlarge the retrieved answer set,
and therefore is a principled way to recover missing reference answers and improve relative completeness.
The trade-off is that fewer constraints may also introduce more spurious answers and larger intermediate results.
}

\revise{
We drop only \textit{one} constraint at a time to enumerate \textit{minimal} relaxations, which has three benefits:
(i) it preserves selectivity as much as possible by avoiding overly aggressive relaxation,
(ii) it helps isolate which constraint is responsible for blocking coverage, and
(iii) it keeps the branching factor controllable under a fixed beam budget.
Descendants are prioritized by Eq.~\ref{eq:score} and only the top-$b$ descendants are expanded to the next beam step.
This favors relaxing less reliable and more permissive constraints via uncertainty $u$,
while also accounting for the parent query's quality score.
}

\revise{
For example, in the demonstration in Appendix~\ref{app:demo}, the initial candidate $Q_0$ includes two season constraints (e.g., \texttt{e.type='Winter'}
and \texttt{e.type='Summer'}) and is over-restrictive and fails the completeness check against $A$.
Dropping the season constraint yields a season-agnostic query $Q_U$, which no longer relies on the uncertain enum grounding and therefore recovers coverage of $A$.
This illustrates why single-constraint relaxation is effective: it makes progress toward coverage with the smallest
possible relaxation at each step, while avoiding unnecessary over-relaxation (and cost) that would occur if multiple
constraints were dropped simultaneously. Finally, \qbc can re-introduce a \textit{KG-grounded} season constraint to
remove spurious answers while preserving coverage, producing the minimal query $Q_M$.
}

\section{\three{Demonstrations}}
\label{app:demo}

\begin{example}
\label{exa-cab}
Considering the NL query {\em ``Which cities in the USA have hosted the Olympics in February?''}, With \begin{lstlisting}[language=SQL, basicstyle=\ttfamily\footnotesize]
A = {Lake Placid; Salt Lake City; Squaw Valley}
\end{lstlisting}, \qc\ produces $Q_U$ retrieving both Summer/Winter hosts, ensuring $G(A)\subseteq Q_U(G)$:
\begin{lstlisting}[language=SQL, basicstyle=\ttfamily\footnotesize]
MATCH (c:City)-[:hosted]->(e:OlympicGames)
WHERE c.country = 'USA' AND e.year > 1896
AND (e.type = 'Summer' OR e.type ='Winter') 
RETURN c.name
\end{lstlisting}
\qbc then adds back only the necessary constraints. 
As February is a winter month,
and the first Olympics hosted in the USA was in 1904, the minimized $Q_M$ might be:
\begin{lstlisting}[language=SQL, basicstyle=\ttfamily\footnotesize]
MATCH (c:City)-[:hosted]->(e:OlympicGames)
WHERE e.type = 'Winter' AND c.country = 'USA' 
RETURN c.name
\end{lstlisting}
which excludes summer‑only hosts like \texttt{{St.Louis, Los Angeles, and Atlanta}}, ensuring soundness as $Q_M(G) \subseteq G(A)$, while retaining only the minimal necessary constraints. 
\end{example}

\revise{As an in-depth illustration of the above example, 
we provide a step-by-step illustration of the \cc pipeline, as a real-world knowledge search use case.}

\stitle{M1: Query-centric Factual Extraction:}

\textbf{Entity Linking}: Initially, entities explicitly mentioned in the query are identified and linked to corresponding nodes in the knowledge graph ($KG$). For our example query, the identified entities are:

\begin{lstlisting}[language=SQL, basicstyle=\ttfamily\footnotesize]
{OlympicGames, USA}
\end{lstlisting}

\textbf{Subgraph Extraction}: Using the linked entities, a relevant subgraph $G'$ is extracted from the main knowledge graph $G$. Typically, neighbors within $k$-hop (usually $k=1$ or $k=2$) from linked entities are included. For simplicity, we consider $1$-hop neighbors, resulting in factual constraints (triples) for the initial \ctable:

\begin{table}[H]
\centering
\begin{tabular}{l|l|l}
?city & hosted  & ?e:OlympicGames \\
\hline
?city & country & USA          \\
\hline
?city & name    & ?city\_name  \\
\hline
?e:OlympicGames & type & ?season   
 \\
\hline
?e:OlympicGames & year & ?host\_year    
\end{tabular}
\end{table}

Here, variables act as placeholders for entities or attributes that will become nodes or properties in the generated Cypher query:

\bi
\item \texttt{?city} represents nodes of type \textit{City}.

\item \texttt{?city$\_$name} denotes the name attribute of a city.

\item \texttt{?season} indicates the type of Olympic Games (e.g., Summer or Winter).

\item \texttt{?host$\_$year} represents the year in which the Olympics were hosted.
\ei

\stitle{M2: CTable Construction (\agent $a_1$):}


Agent $a_1$ systematically extracts relevant triples from the subgraph and enriches these factual constraints with implicit conditions derived from its internal knowledge base. Specifically, $a_1$ extends the constraints in \ctable based on contextual relevance to the query:

\begin{table}[H]
\centering
\begin{tabular}{c|l}
ID & Constraint in \ctable \\
\hline
$c_1$ & \texttt{(c:City)-[:hosted]->(e:OlympicGames)} \\
$c_2$ & \texttt{c.country = 'USA'} \\
$c_3$ & \texttt{e.year > 1896} \\
$c_4$ & \texttt{e.type = 'Winter'} \;\; \\
$c_5$ & \texttt{e.type = 'Summer'} \;\; 
\end{tabular}
\end{table}


Here, additional constraints include:

\bi
\item \texttt{e.year $>$ 1896} ensures relevant modern Olympic events.

\item \texttt{e.type within [Summer, Winter]} are temporal constraints, restricting the type of Olympic events to major recognized categories. For this query, the temporal cue \emph{``in February''} should ideally map to a winter-specific constraint, but in practice, the LLM may \emph{fail to ground} the correct season predicate/value under a given KG schema.

\ei






We compute an uncertainty score $u(c)$ based on capped match counts in $G$
(see Sec.~\ref{sec:method}-B). Intuitively, constraints that are (i) very broad or (ii) unreliable to ground
tend to be deprioritized during search.

\stitle{M3: Chase \& Backchase (\agent $a_2$):}

\agent $a_2$ first retrieves the reference answer set $A$ from an \llm oracle, which provides a reference answer set $A$:

\begin{lstlisting}[language=SQL, basicstyle=\ttfamily\footnotesize]
A = {Lake Placid; Salt Lake City; Squaw Valley}
\end{lstlisting}

\textbf{\qc (Universal query generation).}
\qc aims to produce a universal query $Q_U$ such that $KG(A) \subseteq Q_U(G)$.
It starts from an initially tight constraint set and iteratively relaxes it by dropping \emph{one} constraint
at a time (Alg.~\ref{alg:chase}, ln.~10), keeping only the top-$b$ candidates per iteration.

Starting from $\C_0=\{c_1,c_2,c_3,c_4,c_5\}$, the generated query enforces both seasonal filters
(\texttt{Winter} and \texttt{Summer}), which is over-restrictive and may return empty results:

\begin{lstlisting}[language=SQL, basicstyle=\ttfamily\footnotesize]
Q_0:
MATCH (c:City)-[:hosted]->(e:OlympicGames)
WHERE c.country='USA' AND e.year>1896
  AND e.type='Winter' AND e.type='Summer'
RETURN c.name
\end{lstlisting}

$Q_0$ fails the completeness check against $A$. 
Assuming beam width = 1, then \textbf{\texttt{e.type='Winter'}} will be dropped in round 1 because it is more specific, given there are fewer winter Olympics, while $Q_1$ still fails the completeness check against $A$. 
\textbf{\texttt{e.type='Summer'}} dropped in round 2, then the $Q_2$ is independent of sessions, which can be written as:


\begin{lstlisting}[language=SQL, basicstyle=\ttfamily\footnotesize]
Q_U:
MATCH (c:City)-[:hosted]->(e:OlympicGames)
WHERE c.country = 'USA'
AND e.year > 1896
RETURN c.name
\end{lstlisting}

This query retrieves all US cities hosting the Olympics after 1896, covering both Summer and Winter games, thus ensuring $KG(A) \subseteq Q_U(KG)$:

\begin{lstlisting}[language=SQL, basicstyle=\ttfamily\footnotesize]
{St.Louis;Los Angeles;Lake Placid;Atlanta;Salt Lake City;Squaw Valley}
\end{lstlisting}

\textbf{\qbc (Minimal Query Refinement)}: $a_2$ then construct a minimal query $Q_M$, ensuring soundness and precision while eliminating irrelevant results:

\begin{lstlisting}[language=SQL, basicstyle=\ttfamily\footnotesize]
MATCH (c:City)-[:hosted]->(e:OlympicGames)
WHERE e.type = 'Winter'
AND c.country = 'USA' 
RETURN c.name
\end{lstlisting}

This minimal query specifically targets cities hosting Winter Olympics (as it queried Feb sessions), thereby excluding irrelevant summer-hosting cities like \texttt{{St.Louis, Los Angeles, and Atlanta}}, satisfying the soundness criterion $Q_M(KG) \subseteq KG(A)$.

\noindent\revise{
\stitle{Error propagation under an incorrect oracle reference set.}
\uqg’s completeness/soundness guarantees are \emph{relative} to the oracle-provided reference set $A$.
To illustrate how Oracle errors propagate, consider the same NL query in this demonstration.
Suppose the correct reference answers are winter hosts
$A=\{\text{Lake Placid},\text{Salt Lake City},\text{Squaw Valley}\}$,
but the oracle mistakenly adds a summer host (false positive), e.g.,
$A^{\mathrm{err}} = A \cup \{\text{Los Angeles}\}$.
In this case, \qc must construct a universal query $Q_U$ whose answers cover $G(A^{\mathrm{err}})$, which forces $Q_U$
to relax or drop the winter-only constraint (otherwise it cannot cover \text{Los Angeles}).
Then \qbc tightens the query while preserving coverage of $A^{\mathrm{err}}$; as a result, the returned minimal query may
retain a summer-permitting season filter (or become season-agnostic), producing answers that deviate from the intended
“February/Winter” semantics.
This demonstrates that if $A$ is wrong, \uqg may overfit to $A$ even though its runtime checks remain valid w.r.t. $A$;
we therefore recommend obtaining $A$ via higher-trust sources such as gold queries, and possible optimization like multi-prompt consensus and external fact checking.
}

\section{Prompt Templates}
\label{app:prompt}

\subsection{\con}
I am generating Cypher queries for natural language (NL) question based on \{KB\}. The first step is to generate a constraint table for a given natural language question and extract entities with their {k}-hop relations. The table should include basic triples in the format (subject, predicate, object) or (instance, filter, value), using \{KB\} entity IDs where applicable.
Please provide as many details as possible, listing all possible triples we might use. Include any relevant relationships, properties, and filters that could be useful in constructing the query. Alao feel free to add any relevant triples or constraints based on your knowledge of {KB} that might improve the generated Cypher query.

\sstab
Requirements:

1. Please only output the table in markdown format, no other context.

2. Double-check that all essential properties for connecting entities are included; don't omit them.

...

\sstab
Input:

\{NL Query\}, 

\{Extracted Entities\}, 

\{Factual Constraints\}

\sstab
Output: 

\{constraint table\}

\subsection{\gen}

Given a natural language query, extracted entities, and a list of constraints, generate a Cypher query that uses all provided constraints without adding additional ones. The query should be optimized for performance testing with the specified constraint set. Add operators or query modifiers as needed to accurately answer the NL query.

**Please be very strict with the syntax rules and requirements mentioned below!!!**

Syntax Rules:

1. For node IDs, use the format: ({`~id`: "<url>"})

2. For constrains/relationships(predicates that connect two nodes) use the format: [:`<url>`]
3. For properties/attributes, use the format m.`<url>`

...

Requirements:

1. Incorporates all given constraints exactly as provided

2. Uses the known entities appropriately

3. Does not introduce any additional constraints

4. Includes necessary operators or query modifiers to answer the NL query

...

\sstab
Input:

\{NL Query\}, 

\{Extracted Entities\}, 

\{Constraints\}

\sstab
Output: 

\{Cypher query\}

\subsection{\eva}

You are an expert system for evaluating answers to knowledge graph queries. Your task is to score a list of potential answers to a given natural language query. Also, output a reference answer set, and then we can assert the quality with self-defined hard constraints.

Please follow these guidelines:

1. Consider the relevance and accuracy of each answer in relation to the query.

2. Assign a score from 0 to 20 for each answer, where:

   * 0 means completely irrelevant or incorrect
   
   * 20 means highly relevant and likely correct
   
   * Use the full range of scores to differentiate between answers
   
3. For empty or null answers, assign a score of 0.

4. If multiple answers are identical, give them the same score.

5. Consider both the content and the format of the answer.

...

\sstab
Input format: 

\{Natural language query\}

\{List of answers\}

\sstab
Output: 

\{Reference Answer Set\}

\{List of scores\}

\revise{
\subsection{Prompt variance of the oracle reference set $A$}
\label{app:oracle-prompt}
}

UniQGen uses an LLM oracle to obtain a reference answer set $A$ for an NL query, which guides the Chase/Backchase loop.

\stitle{P0: Minimal answer-only prompt (fast, higher variance).}
\begin{quote}\small
\textbf{Prompt:} Given the question: \emph{``\{NL query\}''}, output the answer entities as a semicolon-separated list.
Do not provide any explanation.
\end{quote}

Note: This prompt is lightweight and cost-efficient, but often exhibits higher variance and may include both false positives and false negatives.

\stitle{P1: Structured, precision-oriented prompt (safer grounding).}
\begin{quote}\small
\textbf{Prompt:} Given the question: \emph{``\{NL query\}''}, output a semicolon-separated list of answer entities.
\textbf{Rules:} (i) only output entities you are confident exist in the KG; (ii) if uncertain, output \texttt{UNKNOWN};
(iii) do not include explanations or extra entities.
\end{quote}

Note: This strategy aims to reduce hallucinated entities (false positives), which helps avoid forcing QChase to relax constraints to cover incorrect answers.

\stitle{P2: Self-consistency oracle (variance reduction via sampling).}
We run the structured prompt (P1) $K$ times with different random seeds (e.g., $K=3$) to obtain $\{A^{(1)},\dots,A^{(K)}\}$, and then aggregate:
\bi
\item \textbf{Majority vote:} include an entity if it appears in at least $\lceil K/2 \rceil$ runs (balanced precision/recall).
\item \textbf{Intersection:} include only entities appearing in all runs (high precision, potentially lower recall).
\ei

Note: Self-consistency often stabilizes $A$ and reduces prompt sensitivity, at the cost of additional oracle calls.

\stitle{P3: MoE-style aggregation (recall expert + precision expert + verifier).}
We implement a lightweight mixture-of-experts (MoE) style oracle using two complementary prompts and an optional verifier:
\bi
\item \textbf{Recall expert (E\textsubscript{R}):} ``List all plausible answer entities, even if you are not fully sure.''
\item \textbf{Precision expert (E\textsubscript{P}):} ``List only high-confidence answer entities; if unsure, output \texttt{UNKNOWN}.''
\item \textbf{Aggregator:} start from $A_{\cap} = A_{R} \cap A_{P}$; if $A_{\cap}$ is empty, fall back to $A_{\cup}=A_{R}\cup A_{P}$
and optionally apply a verifier prompt:
\begin{quote}\small
\textbf{Verifier:} Given the question and candidate answers, remove any answer you are not confident is correct.
Output the remaining answers only (semicolon-separated).
\end{quote}
\ei

Note:  This MoE-style strategy explicitly trades off recall and precision, mitigating both hallucinated additions and excessive omissions in $A$, while incurring higher token consumption and end-to-end latency.

\section{\revise{Complementary Experimental Study}}
\label{app:exp}

\stitle{Ablation Studies.}
We further analyze critical design choices:

\eetitle{Prompt-only is insufficient.} 
The \blone baseline underperforms on all datasets (e.g., GrailQA-overall F1=$0.306$; GraphQ F1=$0.223$), confirming that prompt engineering alone may not be sufficient to achieve desired performance, while \cc benefits from both prompts 
and \llm-guided constraint-based query generation process.

\noindent\revise{
\eetitle{Evaluator strength matters.}
To explore the effect of LLM Oracle, and potential upper bounds of our evaluation process, we conducted experiments replacing the LLM-generated reference set used in  Chase \& Backchase with groundtruth on $100$ random \gq questions. The direct use of ground truth notably increased F1 from $0.804$ to $0.835$,
showing that a stronger oracle improves \uqg’s acceptance decisions and tightens guarantees without changing the logic of the planner.
This highlights \uqg's ability to offer strong and reliable quality guarantees when guided by a trustworthy evaluator. 
Tests were executed using Claude 3.5:
\begin{table}[H]
\centering
\begin{adjustbox}{max width=0.618\textwidth}
\begin{tabular}{lcccc}
\toprule
\textbf{Setting} & \textbf{P} & \textbf{R} & \textbf{F1} & \textbf{EM} \\
\midrule
w. \llm-based \eva & 0.813 & 0.857 & 0.804 & 0.740 \\
w. Groundtruth Reference Set & 0.833 & 0.898 & 0.835 & 0.800 \\
\bottomrule
\end{tabular}
\end{adjustbox}
\vspace{-2ex}
\end{table}
On the other hand, we evaluate oracle stability by measuring model–ground-truth agreement on a random sample of 200 questions. The oracle achieves an average overlap rate above 80\%, indicating that, given the current capability of LLMs, using an LLM oracle provides a sufficiently reliable signal to support our framework in practice.
}

\noindent\revise{
\stitle{GraphQ IR v.s. \uqg.} GraphQ IR~\cite{nie2022graphq} that outputs a unified IR and compiles it to multiple query languages, which is different from \uqg’s rendering-based, runtime-validated approach. 
We evaluate GraphQ IR on the overlapping benchmark it supports (GrailQA), and report the accuracy in the following table:
\begin{table}[H]
\centering
\begin{adjustbox}{max width=0.618\textwidth}
\begin{tabular}{lcccc}
\toprule
\textbf{Methods} & \textbf{I.I.D} & \textbf{Compositional} & \textbf{Zero-shot} & \textbf{Overall} \\
\midrule
\uqg & 0.989 & 0.74 & 0.889 & 0.871 \\
GraphQ IR & 0.874 & 0.495 & 0.096 & 0.369 \\
\bottomrule
\end{tabular}
\end{adjustbox}
\vspace{-2ex}
\end{table}
It is clear that \uqg outperforms GraphQ IR across all splits of the GrailQA benchmark. For GraphQ and WebQSP, GraphQ IR does not include an official setting in its release. Adapting it to these benchmarks would require creating IR annotations (or an equivalent supervision signal) and training a new parser based on their self-defined IR formalism. Therefore, we do not present GraphQ IR results for GraphQ/WebQSP.
}

\stitle{Case Studies.}
Benchmarks like \webqsp often reward literal truth-matched retrieval, overlooking user intent.  
For example, WebQTest-1572 asks: ``{\em what should I do \textbf{today} in San Francisco?}''  
Its ``golden'' SPARQL simply lists all tourist attractions linked to the city.  
When we tested this on a Wednesday, Agent~$a_1$ automatically added constraints that removed SFMOMA and the Asian Art Museum - both closed on Wednesdays; while the reference set generated by Agent~$a_2$ filtered out the mis-typed business Travefy, a travel-software company mistakenly included in the KG.  
If a customer asked Amazon Alexa this question on a Wednesday, they would expect precisely such context-aware, operational answers.  
This demonstrates how \uqg’s LLM-assisted, constraint-based design moves beyond static benchmark evaluation toward deployable, intent-aligned query generation.

\begin{table}[t]
\centering
\small
\setlength{\tabcolsep}{4pt}
\begin{tabular}{l|l|c|c|l}
\hline
\textbf{Methods} & \textbf{Category} & \textbf{Train} & \textbf{Query} & \textbf{WebQSP Performance} \\
\hline
CBR-KBQA (CBR)~\cite{das2021case} & Case-based reasoning (non-parametric) & Y & Y & F1=72.8, EM=70.0 \\
QGG~\cite{lan2020query} & Semantic parsing (step-wise) & Y & Y & F1=74.0 \\
RNG-KBQA~\cite{ye2022rng} & Semantic parsing (seq2seq) & Y & Y & F1=75.6, EM=71.1 \\
NSM~\cite{he2021improving} & Neural symbolic / GNN & Y & N & F1=74.3 \\
UniKGQA~\cite{jiangunikgqa} & GNN-based retrieval & Y & N & F1=72.2, Hit@1=77.2 \\
StructGPT~\cite{jiang2023structgpt} & LLM+KG (structured prompting) & N & N & Hit@1=72.6 \\
RoG-7B~\cite{luoreasoning} & KG+LLM (7B, fine-tuned) & Y & N & F1=70.8, Hit@1=85.7 \\
G-Retriever~\cite{he2024g} & GNN+LLM & Y & N & Hit@1=73.79 \\
\hline
\end{tabular}
\caption{\revise{Representative KGQA methods and reported performance on WebQSP. “Train” indicates task-specific training; “Query” indicates whether the method outputs an executable query (e.g., SPARQL/logical form) rather than only answers. Reported metrics vary across papers (F1/EM vs Hit@1).}}
\label{tab:webqsp_baselines}
\end{table}

\noindent\revise{
\stitle{Representative KGQA methods.}
Table~\ref{tab:webqsp_baselines} summarizes representative KGQA approaches on WebQSP spanning
non-parametric case-based reasoning (CBR-KBQA), classical semantic parsing (QGG, RNG-KBQA),
neural-symbolic / GNN-based reasoning (NSM, UniKGQA), and LLM-era KGQA that couples LLMs with graph grounding
(StructGPT, RoG-7B, and G-Retriever).
We highlight two practical dimensions that often lead to different deployment trade-offs:
\textbf{Train} indicates whether a method relies on task-specific training (or a training-backed memory, as in CBR-KBQA),
and \textbf{Query} indicates whether the method outputs an executable structured query (e.g., SPARQL or a logical form)
instead of producing answers directly.
Semantic parsing methods (QGG, RNG-KBQA) and CBR-KBQA explicitly generate or retrieve executable queries and report strong
F1/EM, but typically depend on supervised training and/or candidate generation and schema alignment.
In contrast, GNN/neural-symbolic methods (NSM, UniKGQA) and many LLM-era approaches (StructGPT, RoG-7B, G-Retriever)
often answer via traversal/retrieval/reasoning over subgraphs rather than emitting a final query, leading to different
robustness and efficiency characteristics.
}

\end{document}